\documentstyle[12pt,epsfig]{article}

\renewcommand{\arraystretch}{1.1}

\newcommand{\npb}[3]{Nucl.~Phys.~#1 (19#2) #3}
\newcommand{\prl}[3]{Phys.~Rev.~Lett.~#1 (19#2) #3}

\newcommand{\pr}[3]{Phys.~Rev.~D#1 (19#2) #3}

\newcommand{\lsim}{\raisebox{-0.13cm}{~\shortstack{$<$ \\[-0.07cm] $\sim$}}~}
\newcommand{\gsim}{\raisebox{-0.13cm}{~\shortstack{$>$ \\[-0.07cm] $\sim$}}~}

\newcommand{\ra}{\rightarrow}
\newcommand{\ee}{e^+e^-}
\newcommand{\s}{\\ \vspace*{-3mm} }
\newcommand{\nn}{\noindent}
\newcommand{\non}{\nonumber}
\newcommand{\beq}{\begin{eqnarray}}
\newcommand{\eeq}{\end{eqnarray}}

\newcommand{\tb}{\tan\beta}

\begin{document}

\begin{titlepage}

\begin{flushright}
PM/96--35\\
KA--TP--28--1996\\
December 1996 \\
\end{flushright}

\def\thefootnote{\fnsymbol{footnote}}

\vspace{1cm}

\begin{center}

{\large\sc {\bf The coupling of the lightest SUSY Higgs boson to}}

\vspace{.3cm}

{\large\sc {\bf two photons in the decoupling regime}} 
\vspace{1cm}

{\sc A.~Djouadi$^{1,2}$, 
V. Driesen${}^2$, W. Hollik${}^2$ and J.I.~Illana${}^2$} 

\vspace{1cm}
${}^1$
Laboratoire de Physique Math\'ematique et Th\'eorique, UPRES--A 5032,\\ 
Universit\'e de Montpellier II, F--34095 Montpellier Cedex 5, France.
\\[3mm]
${}^2$
Institut f\"ur Theoretische Physik, Universit\"at Karlsruhe,\\
D--76128 Karlsruhe, FR Germany. \\

\end{center}

\vspace{1.6cm}

\begin{abstract}

We analyze the contribution of the SUSY particles to the coupling of the
lightest Higgs boson to two photons in supersymmetric theories. We discuss to
what extent these contributions can be large enough to allow for a
discrimination between the lightest SUSY and the standard Higgs particles in
the decoupling limit where all other Higgs bosons are very heavy and no
supersymmetric particle has been discovered at future colliders. We find that
only chargino and top squark loops can generate a sizeable difference between
the standard and the SUSY Higgs--photon couplings. For masses above 250 GeV,
the effect of chargino loops on the two--photon width is however smaller than
$\sim 10\%$ in the entire SUSY parameter space. Top squarks heavier than 250
GeV can induce deviations larger than 10\% only if their couplings to the Higgs
boson are large. Since top squark contributions can be sizeable, we derive the
two--loop QCD correction to squark loops and show that they are well under
control. 

\end{abstract}

\end{titlepage}

\def\thefootnote{\arabic{footnote}}
\setcounter{footnote}{0}
\setcounter{page}{2}

\subsection*{1. Introduction}

In supersymmetric (SUSY) theories, the Higgs sector is extended to contain at
least two isodoublets of scalar fields. In the minimal version, the Minimal
Supersymmetric Standard Model (MSSM), this leads to the existence of five Higgs
particles: two CP--even Higgs bosons $h$ and $H$, a CP--odd or pseudoscalar
Higgs boson $A$, and two charged Higgs particles $H^\pm$ \cite{HHG}. Besides
the four masses, two additional parameters are needed to describe the Higgs
sector at tree--level: $\tb$ the ratio of the two vacuum expectation values and
a mixing angle $\alpha$ in the CP--even sector. However, only two of these
parameters are independent, and choosing the pseudoscalar mass $M_A$ and $\tb$
as inputs, the structure of the MSSM Higgs sector is entirely determined. \s 

If the pseudoscalar mass $M_A$ is very large, $M_A^2 \gg M_Z^2$, the pattern of
Higgs masses is quite regular. The heavy CP--even, CP--odd and charged Higgs
bosons are nearly mass degenerate, $M_H \simeq M_{H^\pm} \simeq M_A$, while the
lightest CP--even $h$ particle reaches its maximal mass value. At tree level,
this value is simply a function of $\tb$, $M_h^{\rm max}=M_Z|\cos 2\beta|\leq
M_Z$. However when including the radiative corrections \cite{rad1,rad2}
which grow as the fourth power of the top mass and logarithmically with
the common squark mass, the upper bound is shifted upwards, $M_h^{\rm max}
\simeq 130$ GeV. In this so called decoupling limit \cite{decou}, which in
practice is reached for $M_A \sim 300$ GeV, the lightest SUSY Higgs boson $h$
has almost the same properties as the SM Higgs particle $H^0$ and the MSSM and
SM Higgs sectors look practically the same, with one light Higgs boson with
a mass below $\sim 130$ GeV. \s 

In the case where no genuine SUSY particle and no additional Higgs boson has
been discovered at future high--energy colliders, the task of discriminating
between the lightest SUSY and the standard Higgs bosons, and therefore between
the MSSM and the SM, in the decoupling limit is challenging. Indeed, since both
have almost the same couplings to fermions and vector bosons, the production
rates and the decay branching ratios [when SUSY Higgs decays are kinematically
not allowed] are practically identical. \s 

Only indirectly that one can distinguish between the two models: if the SM is
extended to the GUT scale, the value $m_t \simeq 175$ GeV requires a Higgs
boson heavier than $M_{H^0} \gsim 130$ GeV \cite{vac} in order that the vacuum
remains stable; since in the MSSM, $M_h$ is constrained to be lighter than
$\sim 130$ GeV, the measured Higgs mass will allow to discriminate between the
SM and MSSM scenarios \cite{vac2}. However, one can assume that new physics 
beyond the SM
exists at a scale $\Lambda \lsim 10$ TeV and in this case, Higgs masses in the
range $M_{H^0} \sim M_Z$ will be still allowed. Furthermore, the SUSY Higgs
mass bound $M_h \lsim 130$ GeV is valid only in the MSSM: for more general SUSY
scenarios where the Higgs sector is even more complicated [for instance in the
NMSSM where an additional Higgs singlet is added], the upper bound on the
lightest Higgs mass from triviality can be extended to $M_h^{\rm max}
\sim 150$ GeV \cite{NMSSM}, leaving a room for an overlap between the allowed
$h$ and $H^0$ masses. \s 

A more ``direct" way to discriminate between the standard and the lightest SUSY
Higgs particles is to look at loop induced Higgs boson couplings such as the
$\Phi gg$ \cite{Hgg}, $ \Phi Z\gamma$ \cite{HZp} and $ \Phi \gamma \gamma$
\cite{Hpp} couplings, $\Phi \equiv h$ or $H^0$. In the SM, these couplings are
mediated by heavy quark and $W$ boson loops [only quark loops for the $H^0gg$
coupling]: since their couplings to the Higgs boson grow with the mass, they
balance the decrease of the triangle amplitude with increasing loop mass, and
these particles do not decouple even for masses much larger than $M_{H^0}$. In
supersymmetric theories, additional contributions will be induced by loops with
charged Higgs bosons, charginos and sfermions; Fig.~1. However, since the SUSY
particles do not couple to the Higgs boson proportionally to their masses,
their contributions are expected to be rather small for large masses.
For very heavy SUSY particles, the loop induced vertices
reduce to their SM values and again, no distinction between the SM and the MSSM
can be made. \s 

The $\Phi gg$ vertex can be measured in the main Higgs production process $gg
\ra \Phi$ at hadron colliders, or via the branching ratio BR$(\Phi \ra gg)$
with the Higgs boson produced at $\ee$ colliders. At the LHC, the 
determination of the cross section $ \sigma(gg \ra \Phi)$ to the level of 
ten percent is rather
difficult, due to uncertainties from the QCD corrections [which at
next--to--leading order are very large \cite{ggQCD1,ggQCD2}, increasing the
cross section by almost a factor of two] and to a lesser extent 
from the parton
densities. The branching ratio BR$(\Phi \ra gg)$ is of the order of a few
percent for $M_\Phi \sim 100$ GeV, and its measurement at $\ee$ colliders with
an accuracy of more than a few ten percent is also very difficult due to the
contamination from charm and bottom quarks \cite{HggBR}. \s 

The $\Phi Z\gamma$ vertex can be measured in the decay $Z\ra \Phi
\gamma$ at LEP and SLC if $M_\Phi <M_Z$, or in the reverse decay $\Phi
\ra Z\gamma$ if $M_\Phi >M_Z$ with the Higgs boson produced in the $gg
\ra \Phi$ fusion mechanism at the LHC. However the rates are very small,
BR$(Z \ra \Phi \gamma) \lsim 10^{-6}$ and BR$(\Phi \ra Z \gamma \ra l^+
l^- \gamma) \lsim 10^{-4}$, leading to only a few events at LEP or the LHC and
making the determination of the $\Phi Z \gamma$ vertex with a reasonable
accuracy very difficult. At future $\ee$ colliders with the expected
integrated luminosities $\int {\cal L} \sim  50$ fb$^{-1}$, running a
few months on the $Z$ resonance would allow to obtain a large sample of
$Z \ra \Phi \gamma$ events if $M_\Phi <M_Z$; a precise measurement of
the $\Phi Z \gamma$ coupling would be possible in this case~\cite{13a}. \s 

The prospects for measuring the loop induced $\Phi \gamma \gamma$ vertex
are as follows: \s

$(i)$ At the LHC the production rate for light Higgs bosons is very large,
$\sigma(gg\ra \Phi) \sim 100$ pb \cite{rev}, and despite of the small
branching ratio BR($\Phi \ra \gamma \gamma) \sim 10^{-3}$, one would
still have ${\cal O}(10^{3})$ $\gamma \gamma$ events after filtering out
most of the background events, if the luminosity is high enough, ${\cal
L} \sim 10^{34}$ cm$^{-2}$s$^{-1}$. However, as discussed earlier,
besides the uncertainties from the parton densities, the theoretical
prediction of the production cross sections is affected by large
uncertainties from higher QCD corrections. Since one measures only 
$\sigma \times$BR, a clean extraction of the $\Phi \ra \gamma \gamma$ 
width will be rather difficult. \s 

$(ii)$ At $\ee$ colliders\footnote{At $\ee$ colliders one can also measure the
$\ee \ra \Phi \gamma$ cross section which is built up by loops of heavy
particles \cite{eeHp}; however the cross sections are rather small, and large
luminosities will be required. Another possibility is provided by the process
$\gamma e^- \ra e^- \Phi$ as recently discussed in \cite{14b}.}, the main
production mechanisms for Higgs particles are the bremsstrahlung process $\ee
\ra \Phi Z$ and the $WW$ fusion process $\ee \ra W^* W^* \ra \Phi \bar{\nu}_e
\nu_e$ \cite{rev}. At energies in the range of $\sqrt{s} \sim 500$ GeV, the
cross sections are around 100 fb for each process; even for integrated
luminosities of $\sim 50$ fb$^{-1}$ one would have only a few $\Phi \ra \gamma
\gamma$ events, a sample which does not allow a precise measurement. At higher
energies the cross section for the $WW$ fusion mechanism increases
logarithmically: at $\sqrt{s} \sim 1.5$ TeV and with $\int {\cal L} \sim 200$
fb$^{-1}$, one would have ${\cal O}(100)$ events allowing for a decent
measurement. However, if no SUSY particles have been found at this energy,
their effect on the $\Phi \ra \gamma \gamma$ width will probably be too small
to be visible. \s 

$(iii)$ The most promising way to have access to the $\Phi \gamma \gamma$
coupling is via the single Higgs production in the fusion process $\gamma
\gamma \ra \Phi$ [18--22], with the photons generated by Compton--back
scattering of laser light \cite{XF}. One can tune the energy of the $\gamma
\gamma$ collider such as to produce the Higgs boson as a resonance in the
$s$-channel. If the luminosity of the $\gamma \gamma$ collider is of the same
order as the luminosity of the original $\ee$ collider, large production rates
can be obtained. A measurement of the 
$\Phi \ra \gamma \gamma$ partial decay width with a precision of the order 
of 10\% could be feasible as will be discussed later. \s 

There are several studies of the Higgs--photon coupling in the MSSM
\cite{ppHth} which however mainly focussed on the detectability of the
$h \ra \gamma \gamma$ signal at the LHC. In this paper, we analyze this
coupling with a different perspective: we scan the entire MSSM parameter
space and single out the regions where the SUSY loops could give
significant contributions. Our aim is to answer to the important
question [since the measurement of the Higgs--photon coupling is one of
the most important goals of the presently discussed $\gamma \gamma$
colliders] of how well one needs to measure the $\Phi \ra \gamma \gamma$
width in order to discriminate between the SM and the MSSM Higgs boson
in the decoupling regime, if no SUSY particle has been observed directly
at the LHC or at an $\ee$ collider with a c.m.~energy of $\sqrt{s}=500$
GeV. \s

The paper is organized as follows. In the next section, we describe the Higgs
sector in the decoupling limit and present for completeness the formulae for
the loop contributions to the $\Phi \gamma \gamma $ coupling. In section 3, we
analyze the Higgs production at $\gamma\gamma$ colliders, and estimate the
precision with which the Higgs--photon coupling can be measured. In section 4
we discuss the various contributions and isolate the parameter space in which
these contributions are significant. Our conclusions will be given in section
5. In the Appendix, we derive the QCD correction to the squark loop
contribution to the $\Phi \gamma \gamma$ amplitude.

\subsection*{2. The Higgs--Photon coupling in the MSSM} 

In the MSSM, using $M_A$ and $\tb$ as input parameters, and including the
leading radiative correction which can be parameterized in terms of the
quantity \cite{rad1} 
\beq
\epsilon = \frac{3 G_{F}}{\sqrt{2}\pi^2} \frac{m_t^4}{\sin^2\beta}
\log\left( 1+\frac{m_{\tilde{q}}^2}{m_t^2} \right) 
\eeq
with $m_{\tilde{q}}$ the common squark mass, the CP--even Higgs boson 
masses are given by 
\begin{equation}
M_{h,H}^2 = \frac{1}{2} (M_A^2+M_Z^2+\epsilon) \Bigg[\,  1 \mp 
\sqrt{1 - 4 \frac{ M_A^2 M_Z^2 \cos^2 2\beta + \epsilon(M_A^2 
\sin^2 \beta + M_Z^2 \cos^2 \beta)}{(M_A^2+M_Z^2+\epsilon)^2} }\, 
\Bigg] 
\end{equation}
In the decoupling limit, $M_A^2 \gg M_Z^2$, the Higgs masses approach
the values 
\beq
M_h & \ra & \sqrt{M_Z^2 \cos^2 2\beta +\epsilon \sin^2 \beta}  \non \\
& & \times \left[ 1 + \frac{ \epsilon M_Z^2 \cos^2 \beta }
 {2 M_A^2 ( M_Z^2 \cos^2 2\beta + \epsilon \sin^2 \beta)}  
 - \frac{ M_Z^2 \sin^2 2\beta + \epsilon \cos^2 \beta }{ 2 M_A^2 } \right]
\non \\
M_H &\ra & M_A \left[ 1 + 
 \frac{M_Z^2 \sin^2 2\beta + \epsilon \cos^2 \beta}{2 M_A^2} \right]
\eeq
The $h$ and $H$ boson masses are displayed in Fig.~2 as a function of
the pseudoscalar mass for several values of $\tb= 1.1, \, 1.6, \, 5$ and
$50$ and for $m_{\tilde{q}}$=250 GeV and 1 TeV. In the case of $h$, the 
decoupling limit $M_h \simeq M_h^{\rm max}$ is reached very quickly for 
large values of $\tb$ [already for $M_A \sim 110$ GeV] and the maximal $h$ 
mass is large, up to $M_h^{\rm max} \simeq 130$ GeV. For small $\tb$ values, 
the maximum $h$ mass is rather small  for $\tb=1.1$; this is due to the
fact that $\cos 2 \beta$ is close to zero and $M_h$ is entirely generated
through radiative corrections. The approach to the decoupling limit is
rather slow, and for $\tb=1.6$, the value $M_h^{\rm max} \simeq 80$--100 GeV
is reached only for $M_A \simeq 500$ GeV. In the decoupling limit, the
heavy CP--even Higgs particle becomes degenerate with the pseudoscalar,
$M_H \sim M_A$. Similarly to $h$, this occurs very quickly for high 
$\tb$ and slowly for low $\tb$ values. \s 

The charged Higgs boson mass is not affected by the large radiative 
correction eq.~(1) and does not depend on $\tb$, it is simply given by 
\beq
M_{H^\pm} &=& M_A \left[ 1+ \frac{M_W^2}{M_A^2} \right]^{1/2} \ \ \
\ra M_A \ \ {\rm for} \ \ M_A \gg M_W
\eeq
It is shown in Fig.~2 together with the $h/H$ masses. Finally, the mixing 
angle angle $\alpha$ which also receives large radiative corrections
\beq
\tan 2 \alpha = \tan 2 \beta \frac{ M_A^2+M_Z^2} {M_A^2-M_Z^2+ 
\epsilon/\cos2\beta} \ ; \ \ -\frac{\pi}{2}<\alpha<0 \ ,
\eeq
reaches the values $\alpha \ra \beta-\pi/2$ in the decoupling limit. 
\s

The two--photon decay width of a CP--even Higgs particle $\Phi=h,H$  can be 
written as \cite{HHG}
\beq
\Gamma (\Phi \ra \gamma \gamma) = \frac{G_F \alpha^2 M_{\Phi}^3}{128 
\sqrt{2} \pi^3 } \left| \, \sum_i A_i (\tau_i) \, \right|^2
\eeq
where the scaling variable $\tau_i$ is defined as $\tau_i = M_{\Phi}^2/
4m_i^2$ with $m_i$ the mass of the loop particle. While in the SM one
has only contributions from the $W$ boson and heavy fermions, in the
MSSM additional contributions are provided by the charged Higgs boson,
the two chargino states and the scalar partners of the fermions; Fig.~1.
Factorizing the reduced couplings of these particles to the Higgs boson
and to the photons, the amplitudes of the various contributions read
\cite{HHG}
\beq
A_W &=& g_{\Phi WW} \, F_1 (\tau_W) \non \\
A_f &=& N_c Q_f^2 g_{\Phi ff} \, F_{1/2} (\tau_f) \non \\
A_{H^\pm} &=& g_{\Phi H^+ H^-}\frac{M_W^2}{M_{H^\pm}^2} \, F_0 (\tau_{H^\pm}) 
\non \\
A_{\chi_i} &=& g_{\Phi \chi_i^+ \chi_i^-} \frac{M_W}{m_{\chi_i}} \, F_{1/2} 
(\tau_{\chi_i}) \non \\
A_{\tilde{f_i}} &=& N_c Q_f^2 g_{\Phi \tilde{f}_i \tilde{f}_i } 
\frac{ M_Z^2} {m_{\tilde{f}_i}^2 } \, F_{0} (\tau_{\tilde{f}_i}) 
\eeq
with $N_c$ the color factor and $Q_f$ the electric charge of the (s)fermion 
in units of the proton charge. With the help of the function $f(\tau)$ 
defined by
\begin{equation}
f(\tau) = \left\{ \begin{array}{ll} 
{\rm arcsin}^2 \sqrt{\tau} & \tau \leq 1 \\
-\frac{1}{4} \left[ \log \frac{1 + \sqrt{1-\tau^{-1} } }
{1 - \sqrt{1-\tau^{-1}} } - i \pi \right]^2 \ \ \ & \tau >1 
\end{array} \right. 
\end{equation} 
the spin 1, 1/2 and spin 0 amplitudes are given by \cite{HHG}
\beq
F_1(\tau) &=& [2\tau^2+3 \tau +3(2\tau-1) f(\tau)]/\tau^2 \non \\
F_{1/2}(\tau) &=& - 2 [\tau+(\tau-1)f(\tau)]/\tau^2 \non \\
F_0 (\tau) &=& [\tau -f(\tau)]/\tau^2 
\eeq
The amplitudes are real if the Higgs mass is below the particle threshold,
$M_\Phi < 2m_i$, while they are complex above this threshold. In the limit 
of heavy loop masses, $\tau \ll 1$, these amplitudes reach the asymptotic 
values
\beq
F_{1} \ra +7 \ \ , \ \ F_{1/2} \ra -\frac{4}{3} \ \ {\rm and}
\ \ F_{0} \ra - \frac{1}{3} 
\eeq
Note that while the $W$ and fermion loops give finite contributions in
the asymptotic limit, the contributions of the charged Higgs boson, the
charginos and the sfermions vanish in the large loop mass limit since
the amplitudes $A_i$ are damped by the heavy masses.  \s

\newcommand{\dd}{{\rm d}}
\newcommand{\ba}{\begin{array}}
\newcommand{\ea}{\end{array}}
\newcommand{\pp}{\gamma \gamma}

\subsection*{3. Higgs production at $\gamma\gamma$ Colliders}

The two--photon width of Higgs bosons can be directly measured at 
$\gamma \gamma$ colliders, with the photons generated by Compton
back--scattering of laser beams from electron beams \cite{XF}. The electron
and laser beam polarizations can be chosen such as to tune the photon 
energy spectrum and produce a peak at a fixed energy. The energy
of the $\gamma \gamma$ collider can be as much as $\sim 80\%$ of that 
of the original $\ee$ collider \cite{XB,XF}. 
In the following, we will discuss briefly the production of the SM Higgs 
boson in the intermediate mass range at photon colliders; we will consider
the various backgrounds  and estimate the precision with which the $H^0 \ra 
\gamma \gamma$ width could be measured. For this purpose we will follow 
Ref.~\cite{XB} from which the  collider configuration, like beam polarization, 
conversion distance, {\it etc}., is taken from. We will assume that the 
Higgs boson mass $M_H$ is already known and therefore, choose the option 
in which the beam energy is tuned for the $\pp$ luminosity to peak at $M_H$. 
We also adopt the option where the handedness of the electron/positron 
beams and laser photons are opposite in order to enhance the $J_Z=0$ 
partial wave in which the Higgs boson signal occurs. \s

For masses below $M_H \lsim 130$ GeV, the Higgs boson will
dominantly decay into $b\bar{b}$ pairs; decays into charm quarks, $\tau$
leptons and gluons occur at the level of a few percent and for masses
close to $M_H \sim 130$ GeV, the $WW$ decay mode becomes important
and reaches a branching ratio of $\sim 30\%$ \cite{decay}. The Higgs
boson is extremely narrow, with a total decay width below $\sim 10$ MeV.
Being induced by loops, the $H^0 \ra \gamma \gamma$ width is very
small, of order of a few 10 keV in the mass range 80 GeV $\lsim M_H
\lsim$ 130 GeV. For a complete discussion, see Ref.~\cite{decay} from 
which we take the inputs for masses and couplings. \s
 
Since the process we are interested in proceeds through a very narrow 
resonance, the detector accuracy when comparing the $\pp \ra H^0 \ra b\bar{b}$ 
signal and the background should be taken into account. A simple way to 
obtain the {\em effective} signal and backgrounds consists of introducing a 
gaussian smearing of the two--photon invariant mass $W$,
\beq
L_{\rm eff}\ \frac{\dd\sigma^{\rm eff}}{\dd W} (W) = 
\int^{y_m\sqrt{s_{e^+e^-}}}_{M_X} \dd W' 
\frac{1}{\sqrt{2\pi}\delta}\exp\left\{-\frac{(W'-W)^2}{2\delta^2}\right\}
\frac{\dd L}{\dd W'} \ \hat{\sigma}(W')
\label{eff}
\eeq
and selecting events within a bin of invariant masses $M_H \pm \Delta$. 
In the previous expression, $L_{\rm eff}$ and $y_m \sqrt{s_{\ee}}$ are
the effective luminosity and the maximum energy of the $\pp$ collider;
$\delta$ is one sigma of the detector resolution for $W$. 
The cross section for the signal process $\gamma\gamma\to H^0 \to b\bar{b}$ 
can be written as
\beq
\hat{\sigma}_{\rm SG}(W) = 4\pi^2\frac{\Gamma(H^0 \to\gamma\gamma)
{\rm BR}(H^0 \to b\bar{b})}{M^2_H} (1+\lambda_1\lambda_2)\delta(W-M_H)\ ,
\label{sg}
\eeq
where the helicities of the scattered photons must be such that $\lambda_1 
\lambda_2=1$. Inserting the cross section in eq.~(11), and selecting
the events in the bin $M_H \pm \Delta$, one obtains
\beq
L_{\rm eff}\ \sigma^{\rm eff}_{\rm SG} (M_H) = R(\Delta/\delta) \
\left.\frac{\dd L}{\dd W}^{J_Z=0}\right|_{W=M_H} \
8\pi^2\frac{\Gamma(H^0 \to\gamma\gamma)B(H^0 \to X)}{M^2_H}\ ,
\label{signal}
\eeq
with $R(\Delta/\delta)$ being the Gaussian error function, describing the
fraction of signal events contained in the bin $M_H\pm\Delta$ 
[for instance, for $\Delta=1.25\delta$ one has $R \simeq 0.75$]. \s

In the intermediate mass range, 80 GeV $\lsim M_H\lsim 130$ GeV, the 
main source of background is the continuum production of $b$-- and 
$c$--quark pairs, including gluon radiation which leads to fake two--jet 
events \cite{XD}. The contribution of resolved photons to heavy quark 
production is rather controversial since the partonic distribution 
functions for polarised photons are not yet available; see the discussions 
given in \cite{XB,XE}. 
Another potentially large background for $M_H \sim M_Z$, is the process 
$e \gamma \to (e)Z \ra b\bar{b}$ which comes from the residual electrons 
that were left over from the Compton scattering; the scattered electron 
is emitted backwards down the beampipe. It can be reduced 
by removing the residual electrons from the interaction region with a 
strong magnetic field, and this requires a non zero conversion distance
\cite{XF}.  Finally, we have  $\gamma\gamma\to Z(f\bar{f}) \ra b \bar{b}$ 
\cite{XB,XG} which would constitute a serious problem that may 
be overcome using a very peaked $\gamma\gamma$ luminosity distribution at 
$W \approx M_ Z$ to greatly increase the signal--to--background ratio.
\s

Here, we will only include the continuum $q\bar{q}$ and $q\bar{q}(g)$ 
backgrounds that we calculated using the package {\tt HELAS} \cite{XH}.
As in Ref.~\cite{XC}, we have used the following 
set of experimental cuts for the signal and backgrounds: 
(i) both jets from the $q$ and $\bar{q}$ 
should be visible in the detector: $|\cos\theta_{q,\bar{q}}|<0.7$; 
(ii) the gluon jet should escape detection: $|\cos\theta_g| > 0.9$;
(iii) the jets should be clearly isolated: $m^2_{ij}/{s_{e^+e^-}}>0.02$;
(iv) the missing $p_T$ and the aplanarity due to missing gluon
should be small: $\not{p_T}<10$~GeV and $||\phi_{q}+\phi_{\bar{q}}|-\pi|<
0.02$. For the detector accuracy, we also employ the same resolution
as in Ref.~\cite{XC}: $\delta=4$ GeV and $\Delta=5$ GeV for half of 
the width of the selection interval. The effective cross sections 
for a tuned energy $0.8 \sqrt{s_{\ee}}=M_H$ are given in Table~1
for three choices of the Higgs boson mass $M_H=80,105$ and 130 GeV. \s

\renewcommand{\arraystretch}{1.2}

\begin{center}
\begin{tabular}{|l|c|c|c|}
\hline
Process & $M_H=80$ GeV & $M_H=105$ GeV & $M_H=130$ GeV \\
\hline\hline
$\gamma\gamma\to H^0 \to b\bar{b}$ & 67.9 & 73.0 & 62.6 \\
\hline\hline
$\gamma\gamma\to b\bar{b}$      & 18.1 & 7.13 & 3.40 \\
\hline
$\gamma\gamma\to c\bar{c}$      & 240  & 99.4 & 49.2 \\
\hline
$\gamma\gamma\to b\bar{b}(g)$   & 0.13 & 0.05 & 0.02 \\
\hline
$\gamma\gamma\to c\bar{c}(g)$   & 1.72 & 0.72 & 0.36 \\
\hline
\end{tabular}
\end{center}
\noindent {\small Table~1: Effective cross sections [in fb] for the signal 
and the backgrounds for a tuned energy $0.8 \protect \sqrt{s_{e^+ e^-} }=M_H$ 
and the luminosity distribution of Fig.~16 of Ref.~\protect\cite{XB}.}

\smallskip
\renewcommand{\arraystretch}{1}

For the $b\bar{b}$ final state, we will assume a detection efficiency 
of 50\% [which should be achieved in the future by micro--vertex detectors] 
with a 5\% contamination from $c\bar{c}$ final states. One the other hand, 
only the events where both quarks decay hadronically should be collected
to estimate the $\gamma\gamma$ invariant mass; the hadronic decay branching 
ratios of $b$-- and $c$--flavored hadrons are 75\% and 82\% respectively. 
Multiplying the tagging efficiencies times the square of the hadronic
branching ratios, we obtain the corrected effective cross sections in 
Table~2. As can be seen, the radiative background is completely negligible
and the signal cross sections are much larger than the backgrounds, especially
for high Higgs boson masses, leading to a large statistical significance 
for the Higgs boson signal. \s

\renewcommand{\arraystretch}{1.2}

\begin{center}
\begin{tabular}{|l|c|c|c|}
\hline
Process & $M_H=80$ GeV & $M_H=105$ GeV & $M_H=130$ GeV \\
\hline\hline
$\gamma\gamma\to H^0\to b\bar{b}$ & 19.1  & 20.5  & 17.6 \\
\hline\hline
$\gamma\gamma\to b\bar{b}$      & 5.09  & 2.00  & 0.96 \\
\hline
$\gamma\gamma\to c\bar{c}$      & 8.07  & 3.34  & 1.65 \\
\hline
$\gamma\gamma\to b\bar{b}(g)$   & 0.037 & 0.014 & 0.006 \\
\hline
$\gamma\gamma\to c\bar{c}(g)$   & 0.058 & 0.024 & 0.012 \\
\hline\hline
Total Background & 13.3 & 5.38 & 2.63 \\
\hline\hline
Signal/Background & 1.44 & 3.81 & 6.69 \\
\hline
Stat. Significance & 16.6 & 27.9 & 34.3 \\
\hline
Sensitivity to $\Gamma_{\gamma\gamma}$ & 9.4$\%$ & 7.8 $\%$ & 8.1 $\%$ \\
\hline
\end{tabular}
\end{center}
\noindent {\small Table 2: Corrected effective cross sections [in fb] to 
include tagging efficiency and the invariant mass reconstruction. 
An integrated luminosity of 10 fb$^{-1}$ is assumed for the evaluation of 
the statistical significance and the sensitivity to the two--photon width 
of the Higgs boson.}
\smallskip
\renewcommand{\arraystretch}{1}

The measurement of $\Gamma(H^0 \to\gamma\gamma)\times {\rm 
BR}(H^0\to\gamma\gamma)$ 
will follow from eq.~(\ref{signal}) if the luminosity, the tagging and 
mass reconstruction efficiencies as well as the Higgs boson mass are
precisely known. Assuming that BR$(H^0 \to\gamma\gamma)$ is given by the SM
and that the uncertainties in all the previous quantities are negligible,
the statistical error in the $\Gamma(H^0 \to\gamma\gamma)$ determination is
$$
\frac{\Delta\Gamma}{\Gamma}=\frac{1}{\sqrt{L_{\rm eff}}}\frac{\sqrt{S+B}}{S}
$$
which is about 10\% for an effective luminosity $L_{\rm eff}=$ 10 fb$^{-1}$.
Increasing the luminosity and improving the $b$--tagging efficiency and
purity as well as the reconstruction of the $b\bar{b}$ invariant mass would
enhance the sensitivity to the $h\gamma\gamma$ coupling.  

\subsection*{4. Loop contributions in the MSSM} 

\subsubsection*{4.1 $W$ boson loop}

Compared to the SM case where $g_{H^0WW}=1$, the $W$ boson amplitude for 
the lightest MSSM Higgs particle $h$ is suppressed by a factor $g_{h WW} 
=\sin(\beta -\alpha)$. However, in the decoupling regime $M_A^2 \gg M_Z^2$, 
the $hWW$ coupling approaches quickly the SM coupling
\beq
g_{hWW} \ = \ \sin(\beta-\alpha) \ \ra \ 1- \frac{1}{8} \sin^2 4 \beta 
\, \frac{M_Z^4}{M_A^4} 
 \left[ 1- \frac{\epsilon}{2 M_Z^2 \cos 2\beta} \right]^2
\; \ra \ 1
\eeq

The $W$ boson form factor $A_W$ is shown in Fig.~3a as a function of 
$M_h$ both in the SM and in the MSSM. The MSSM contribution has been 
obtained by fixing the pseudoscalar mass to $M_A =250$ GeV and varying 
the value of $\tb$ from $\tb=1.1$ to $50$. The difference between the SM 
and the MSSM contributions is very small, even for low $\tb$ values where 
the decoupling limit is not completely reached yet for $M_A=250$ GeV. 
This is due to the fact that $\sin(\beta-\alpha)$ approaches unity very
quickly, the difference being of ${\cal O}(M_Z^4/M_A^4)$. 

\subsubsection*{4.2 Fermion loops} 

Since the $\Phi ff$ couplings are proportional to the fermion mass, the
contribution of the light fermions to the $\Phi \gamma \gamma$ amplitude
is negligible. Only the top quark, and to a smaller extent the 
charm and bottom quark, as well as the $\tau$ lepton, 
will effectively contribute. Compared to the SM case where
$g_{H^0ff}=1$, the $huu/hdd$ couplings are suppressed/enhanced by the
factors 
\beq
g_{huu} &=& \ \ \frac{\cos \alpha}{\sin \beta}  \  \ra \ 
1 + \frac{1}{2} \frac{M_Z^2}{M_A^2} \cot \beta \sin 4\beta \; 
 \left[ 1- \frac{\epsilon}{2 M_Z^2 \cos 2\beta} \right]
\ \ra 1 \non \\
g_{hdd} &=& -\frac{\sin \alpha}{\cos \beta}  \  \ra \ 
1 - \frac{1}{2} \frac{M_Z^2}{M_A^2} \tb \sin 4\beta  \;
 \left[ 1- \frac{\epsilon}{2 M_Z^2 \cos 2\beta} \right]
\ \ra 1 
\eeq
The fermionic amplitudes $A_t$ and $A_{b,c,\tau}$ are shown in Fig.~3b as a
function of $M_h$, with $M_A$ again fixed to 250 GeV. In the SM, 
the dominant fermionic contribution $A_t$ is almost constant and can be 
approximated by $A_t \sim N_c\, Q_t^2 \, F_{1/2}(0) = -16/9$. 
It is smaller than the $W$ boson
contribution and the two amplitudes interfere destructively. In the 
MSSM, the variation with $M_h$ is rather pronounced. This is due
to the variation of the coupling $g_{huu}$ since the decoupling limit is
not reached yet for $M_A=250$ GeV and small $\tb$ values:
contrary to $g_{hWW}$, the coupling $g_{huu}$ approaches the decoupling 
limit slowly,  $g_{huu} \ra 1-{\cal O} (M_Z^2/M_A^2)$. \s

For the bottom quark loop, the amplitude $A_b$ has both real and
imaginary parts since $M_h >2m_b$. The real part of $A_b$, calculated
with a running mass $m_b(M_\Phi^2) \sim 3$ GeV, is much smaller compared
to $A_t$ as expected\footnote{In the MSSM, however, far from the
decoupling limit and for large values of $\tb$, the amplitude $A_b$ can
be very large since the coupling $g_{hbb} \sim \tb$ is strongly
enhanced.}, but is of the same order as $A_{\tau}$ since the latter
is not penalized by the charge factor.
The imaginary parts Im$(A_{b,\tau})$ are larger than Re($A_{b,\tau}$), but
since they do not interfere with the dominant $A_W$ and $A_t$
contributions, their effect on the $\Phi \ra \gamma \gamma$ width is
rather small. Note that the difference between the SM and MSSM is still
rather large for $M_A=250$ GeV, however, this difference will hardly be
noticed in $\Gamma (h \ra \gamma \gamma$) since the contributions of the 
$b$ and $\tau$ loops are small. \s

Finally, we note that the QCD corrections to the dominant top quark loop 
are well under control and can be included by simply multiplying the Born 
amplitude by a factor $(1- \alpha_s/\pi)$. The QCD corrections to the $b$ 
quark loop do not exceed the level of a few times $\alpha_s/\pi$ if the 
running $b$ quark mass at a scale $M_\Phi/2$ is used in the Born amplitude; 
for more details on these corrections, see Ref.~\cite{ggQCD2}. 

\subsubsection*{4.3 Charged Higgs boson loops} 

In the coupling of the lightest CP--even Higgs particle $h$ to charged 
Higgs bosons, large radiative corrections which cannot be mapped into 
the mixing angle $\alpha$ will appear. Retaining again only the leading 
correction, the $g_{h H^+ H^-}$ coupling is given by \cite{ghhh}
\beq
g_{h H^+H^+} &=& \sin(\beta-\alpha) +\frac{\cos2 \beta 
\sin(\beta+\alpha)}{2c_W^2} + \frac{\epsilon}
{2c_W^2 M_Z^2} \frac{\cos \alpha \cos^2 \beta}{\sin \beta} 
\eeq
with $s_W^2 =1-c_W^2 \equiv \sin^2\theta_W$. In the decoupling limit, 
the coupling reduces up to ${\cal O}(\epsilon)$ terms, to 
$ g_{h H^+H^+}  \ra 1 - \cos^2 2 \beta/(2c_W^2)$. \s

The form factor $A_{H^\pm}$ is shown in Fig.~4a as a function of
$M_{H^\pm}$ for $\tb=1.6, \, 5$ and $50$. Because the contribution is
damped by a factor $1/M_{H^\pm} ^2$ for large $H^\pm$ masses, and also
because the spin--zero amplitude $F_0$ is small, the charged Higgs
contribution to the $h \gamma \gamma$ coupling is very small. For low
masses, $M_{H^\pm} \sim 100$ GeV, $A_{H^\pm}$ can reach values close to
$\sim -0.1$, but for $M_{H^\pm} \gsim 250$ GeV the contribution of the
$H^\pm$ loop is already only a few per mille of that of the dominant $W$
boson loop, and is therefore completely negligible\footnote{Note that in
two--Higgs doublet models, charged Higgs boson loops will provide the only
additional contribution to the $h \gamma \gamma$ coupling. Since this
contribution is very small, discriminating between this model and the SM
in the decoupling regime using the $h \gamma \gamma$ coupling will not
be possible.}. 

\subsubsection*{4.4 Scalar lepton and quark loops}

The left-- and right--handed scalar partners of each SM charged fermion,
$\tilde{f}_L$ and $\tilde{f}_R$, mix to give the mass eigenstates
$\tilde{f}_1$ and $\tilde{f}_2$. The mixing angle is
proportional to the fermion mass and is therefore important only in the
case of the top squarks \cite{qmix}; for the scalar partners of light
fermions the current eigenstates are identical to the mass
eigenstates\footnote{The mixing in the sbottom sector can also be
sizeable for large values of $\tb$. We have checked explicitely that
this mixing will not affect significantly the numerical results compared
to the no--mixing case, if the value of the off--diagonal entry in the
sbottom mass matrix is not prohibitively large.}. In this subsection, we
will discuss the contribution of slepton and the scalar partners of the
light quarks only, the contributions of top squark loops will be
discussed separately later. \s 

The reduced couplings of the $h$ boson to the left--handed and 
right--handed partners of light fermions, are given by 
\beq
g_{h \tilde{f}_L \tilde{f}_L } &=& (I_3^f -Q_f s_W^2) 
\sin(\beta+\alpha) \non \\
g_{h \tilde{f}_R \tilde{f}_R } &=& Q_f s_W^2 \sin(\beta+\alpha) 
\eeq
with $I_3^f =\pm 1/2$ and $Q_f$ the weak isospin and the electric charge
of the fermion $f$. In the decoupling limit, one has $\sin(\beta+ \alpha) 
\ra -\cos 2\beta$. \s

The contributions of the slepton and squark [except for top squark] 
loops are shown in Fig.~4b as functions of the masses and for the three
values $\tb=1.6, \, 5$ and 50 with $M_A$ fixed to 250 GeV. We have
summed over all slepton and squark [except stop] contributions, and used 
common masses $m_{\tilde{l}}$ and $m_{\tilde{q}}$, which is approximately
the case in SUSY--GUT models. The form factor $A_{\tilde{l}}$ is
approximately equal to $A_{H^\pm}$ except that the trend for various
$\tb$ values is reversed. The contribution of the squark loops
$A_{\tilde{q}}$ has almost the same magnitude as the contribution
$A_{\tilde{l}}$, but is of opposite sign. As in the case of the charged
Higgs boson, slepton and squark loop contributions to the $h \ra \gamma
\gamma$ decay width are very small: in the decoupling limit and for
loop masses above 250 GeV, they do not exceed a few per mille and can be 
safely neglected. 

\subsubsection*{4.5 Top squark loops}

Due to the large value of the top quark mass, the mixing between the 
left-- and right--handed scalar partners of the top quark, $\tilde{t}_L$ 
and $\tilde{t}_R$, can be very large. The mass eigenstates $\tilde{t}_1$ 
and $\tilde{t}_2$ are obtained by diagonalizing the mass matrix
\begin{equation}
{\cal M}^2_{\tilde{t}} = \left(
  \begin{array}{cc} m_{\tilde{t}_L}^2 + m_t^2 + \cos 2 \beta 
(\frac{1}{2}    - \frac{2}{3}s_W^2) \, M_Z^2  & m_t \, m^{LR}_t \\
                    m_t m^{LR}_t & m_{\tilde{t}_R}^2 + m_t^2
                   + \frac{2}{3}\cos 2 \beta \; s_W^2 \, M_Z^2
\end{array} \right)
\end{equation}
where the left-- and right--handed scalar masses $m_{\tilde{t}_L}$ and 
$m_{\tilde{t}_R}$ are generally assumed to be approximately equal to the 
common mass of the 
scalar partners of light quarks $m_{\tilde{q}}$. In terms of the 
soft--SUSY breaking trilinear couplings $A_t$ and the Higgs--higgsino 
mass parameter $\mu$, the off--diagonal term $m^{LR}_t$ reads
\beq
m^{LR}_t = A_t - \mu \, \cot \beta 
\eeq
The top squark masses and the mixing angle are then given by
\beq
m^2_{\tilde{t}_1 , \tilde{t}_2} = m_t^2+ \frac{1}{2} \left[ 
m_{\tilde{t}_L}^2 + m_{\tilde{t}_R}^2 \mp \sqrt{( m_{\tilde{t}_L}^2 
-m_{\tilde{t}_R}^2)^2+ (2 m_t m^{LR}_t)^2 } \, \right]
\eeq
\beq
\sin 2\theta_t = \frac{2 m_t m^{LR}_t}{m_{\tilde{t}_1}^2 - 
m_{\tilde{t}_2}^2} \ \ , \ \ \cos 2 \theta_t = \frac{m_{\tilde{t}_L}^2 - 
m_{\tilde{t}_R}^2 } {m_{\tilde{t}_1}^2 - m_{\tilde{t}_2}^2}
\eeq
The couplings of the $h$ boson to top squarks in the 
presence of mixing are given by
\beq
g_{h \tilde{t}_1 \tilde{t}_1 } &=& \frac{1}{2} \sin(\alpha+\beta) 
\left[\cos^2 \theta_t - \frac{4}{3}s_W^2 \cos 2\theta_t \right] - 
\frac{\cos \alpha} {\sin \beta} \frac{m_t^2}{M_Z^2} \non \\
&& + \frac{m_t \sin2\theta_t}{2M_Z^2} \left[ \frac{\cos \alpha}
{\sin \beta} A_t + \frac{\sin\alpha}{\sin\beta}\, \mu \right] 
\non \\
g_{h \tilde{t}_2 \tilde{t}_2 } &=& \frac{1}{2} \frac{\sin(\alpha+\beta)}{2} 
\left[\sin^2 \theta_t + \frac{4}{3}s_W^2 \cos 2\theta_t \right] 
- \frac{\cos \alpha} {\sin \beta} \frac{m_t^2}{M_Z^2} \non \\
&& - \frac{ m_t 
\sin2\theta_t} {2M_Z^2 } \left[ \frac{\cos \alpha}{\sin \beta} A_t + 
\frac{\sin\alpha}{\sin\beta}\, \mu \right] 
\eeq
In the decoupling limit, these vertices reduce to 
\beq
g_{h \tilde{t}_1 \tilde{t}_1 } &=& - \frac{1}{2} \cos 2\beta \left[ 
\cos^2 \theta_t - \frac{4}{3} s_W^2 \cos 2\theta_t \right] - 
\frac{m_t^2}{M_Z^2} + \frac{1}{2} \sin 2\theta_t \frac{m_t m^{LR}_t}
{M_Z^2}  \non \\
g_{h \tilde{t}_2 \tilde{t}_2 } &=& - \frac{1}{2} \cos 2\beta \left[ 
\sin^2 \theta_t + \frac{4}{3} s_W^2 \cos 2\theta_t \right] - 
\frac{m_t^2}{M_Z^2} - \frac{1}{2} \sin 2\theta_t \frac{m_t m^{LR}_t}
{M_Z^2}  
\eeq
Assuming as usual that $m_{\tilde{t}_L} = m_{\tilde{t}_R} = m_{\tilde{q}}$,  
the only parameters which enter the contribution of the $\tilde{t}$ loops to
the $h \ra \gamma \gamma$ decay width in the decoupling limit are
$m_{\tilde{q}}$ [that we will trade against $m_{\tilde{t}_1}$] and
$m^{LR}_{t}$. There is also a dependence on $\tb$ which arises from the Higgs
coupling to top squarks and from the mixing angle since the stop mass matrix
contains also a small $\cos 2\beta$ term. However, this dependence on $\tb$ is
rather small. \s

In Fig.~5a, we show contour plots in the $(m^{LR}_t,m_{\tilde{t}_1})$
plane for which the contribution $A_{\tilde{t}}$ [which includes the
amplitudes of both top squarks] to the $h \gamma \gamma$ coupling is
$|A_{\tilde{t}}|= 2,1,0.5$ and $0.2$. For the sake of convenience, we
also display in Fig.~5b, contours in the $(m_t^{LR}, m_{\tilde{t}_1})$
plane for fixed masses of the heavy top squark and the scalar partners
of the light squarks. To have a better insight on the various
contributions, we show in Fig.~6 three dimensional plots of the two
separate top squark amplitudes $A_{\tilde{t}_1}$ and $A_{\tilde{t}_2}$.
\s 

The amplitude $A_{\tilde{t}}$ is symmetric for positive and negative
$m^{LR}_t$ values because for large $m^{LR}_t$, 
the $h \tilde{t} \tilde{t}$ coupling is dominated by the $\sin 2
\theta_t m_t^{LR}$ term and $\sin 2 \theta_t$ is proportional to
$m^{LR}_t$; for small $m_t^{LR}$ the dominant piece of the $h \tilde{t}
\tilde{t}$ coupling is proportional to $m_t^2$. To discuss the effect of
the mixing, it is convenient to divide the parameter space into three
regions: intermediate $|m^{LR}_t|$ values around the region delimited 
by the contour $A_{\tilde{t}}=0$, large and small $|m^{LR}_t|$ values 
away from this contour. \s 

For large $|m^{LR}_t|$, the contributions are large and positive;
for light enough top squarks, $m_{\tilde{t}_1} \sim 100$ GeV, they can
reach the value $A_{\tilde{t}} \sim 2$ for $|m^{LR}_t| \sim 1$ TeV,
therefore almost canceling the top quark loop contribution $A_t$. For a 
given $m_{\tilde{t}_1}$, $A_{\tilde{t}}$ is larger for higher
values of $m^{LR}_t$ because in this case, the coupling $h \tilde{t}
\tilde{t} \sim m^{LR}_t$ is strongly enhanced. For large
$m_{\tilde{t}_1}$, the two top squarks will have comparable masses [see
Fig.~5b], and since the signs in the dominant component of the $h
\tilde{t}_1 \tilde{t}_1$ and $h \tilde{t}_2 \tilde{t}_2$  couplings are
opposite, the two amplitudes will partly cancel each other; Fig.~6. \s 

For small $|m^{LR}_t|$, there is a region [the small ``menhir"
around $m^{LR}_t=0$] where no solution for $m_{\tilde{t}_1} <m_t$ is
allowed when diagonalizing the mass matrix; however, this region is
already excluded by CDF/D0 data from the negative search of scalar
partners of light quarks with masses $m_{\tilde{q}} \lsim 150$ GeV \cite{CDF};
Fig.~5b. The amplitudes in this region are negative since the dominant
component of the $h \tilde{t} \tilde{t}$ is now proportional to $m_t^2$
and has opposite sign compared to the dominant off diagonal coupling 
when $m^{LR}_t$ is large. $A_{\tilde{t}}$ decreases with increasing top 
squark mass as expected, and can reach the almost maximal value 
$A_{\tilde{t}} \sim -0.5$ for $m_{\tilde{t}_1} \lsim 250$ GeV; however, 
most of this region is again ruled out by the CDF/D0 bound $m_{\tilde{q}}
\gsim 150$ GeV as shown in Fig.~5b. \s 

For intermediate values of $|m^{LR}_t|$, there is a balance between
the two components of the $h \tilde{t}_1 \tilde{t}_1$
coupling which tend to cancel each other, and the 
two contributions $A_{\tilde{t}_1}$ and $A_{\tilde{t}_2}$ which become 
comparable [since for relatively large $m_{\tilde{t}_1}$, $\tilde{t}_1$ 
and $\tilde{t}_2$ have comparable masses, Fig.~5b, and $A_{\tilde{t}_1}$ 
is small] and interfere destructively. At some stage, the two contributions 
cancel each other leading to the contour $A_{\tilde{t}} =0$ of Fig.~5a. 
As one can see, for top squarks not much heavier than $\sim 250$ GeV, one 
can have contributions of the order of 10\% or more to the $h \ra \gamma 
\gamma$ decay width if the off--diagonal entry in the stop mass matrix is 
large, $|m^{LR}_t| \gsim 1$ TeV. \s

Since top squark contributions can be relatively large, we have derived 
the two--loop QCD corrections to the scalar quark loops in the limit of 
heavy squarks, using low energy theorems. The derivation of the result 
is done in the Appendix. The effect of the QCD corrections is to shift 
the value of the Born form--factor $A_{\tilde{q}}$ by an amount 
\beq
A_{\tilde{q}} = A_{\tilde{q}}^{\rm Born} \left[1 + \frac{8}{3} 
\frac{\alpha_s} {\pi} \right] 
\eeq
The correction is about three times larger than in the case of quark loops,
and has opposite sign. It is of the order of $\sim 10\%$, and therefore 
well under control. 
\s 

\subsubsection*{4.6 Chargino loops}

The masses of the two chargino states depend on $\tb$, the gaugino mass 
parameter $M_2$ and the Higgs--higgsino mass parameter $\mu$: 
\beq
m_{\chi_{1,2}}^2 &=& \frac{1}{2} \Big[ M_2^2+\mu^2+2 M_W^2 \non\\ 
&&    \mp \sqrt{(M_2^2-\mu^2)^2 +4 M_W^4 \cos^2 2\beta +4 M_W^2 
(M_2^2+\mu^2+2 M_2 \mu \sin 2\beta } \Big]
\eeq
The chargino couplings to the lightest Higgs boson $h$ are given by \cite{SUSY}
\beq
g_{h \chi_1^+ \chi_1^-} &=& \sqrt{2} \left[ \cos \alpha
\cos \theta^+ \sin \theta^- + \sin \alpha \sin \theta^+ \cos 
\theta^- \right] \non \\
g_{h \chi_2^+ \chi_2^-} &=& -\varepsilon \sqrt{2} \left[ \cos 
\alpha \cos \theta^- \sin \theta^+ + \sin \alpha \sin \theta^- \cos 
\theta^+ \right] 
\eeq
with $\varepsilon = {\rm sign}( M_2 \mu - M_W^2 \sin2\beta)$ and the angles
$\theta^\pm$ according to 
\beq
\tan 2 \theta^- &=& \frac{2 \sqrt{2}M_W (M_2 \cos \beta+ \mu \sin \beta)}
{M_2^2 -\mu^2 -2M_W^2 \cos 2\beta} \non \\
\tan 2 \theta^+ &=& \frac{2 \sqrt{2}M_W (M_2 \sin \beta+ \mu \cos \beta)}
{M_2^2 -\mu^2 +2M_W^2 \cos 2\beta} 
\eeq
Note that in the decoupling limit, one has $\cos \alpha=\sin \beta$ 
and $\sin \alpha=-\cos\beta$. \s

The contribution $A_\chi$ of the chargino loops to the $h \gamma \gamma$ 
coupling is shown in Fig.~7 in the $(M_2,\mu)$ plane for the two
values $\tb=1.6$ and 50. Contours for $|A_\chi| =0.5, 0.3, 0.2$ and $0.1$ as
well as the region of the parameter space for which the lightest
chargino mass is larger than 90 GeV [a value below which the charginos
will be found at LEP2] and 250 GeV [which will be probed at a 500 GeV
$\ee$ collider], have been drawn. \s

The chargino contributions to the $h \gamma \gamma$ coupling are much 
larger than those of charged Higgs bosons, sleptons and the scalar
partners of light quarks. This is due to the fact that for heavy 
particles, the amplitude $F_{1/2} \ra -4/3$ is larger than $F_0\ra 
-1/3$, and also because the chargino contribution scales like $A_\chi 
\sim 1/m_\chi$ contrary to the amplitudes for scalars which scale 
like $A_i \sim 1/m_i^2$. For high $\tb$ the contributions are positive,
while for low $\tb$ the amplitudes follow the sign of $\mu$. \s

The largest contributions are obtained for small values of $\tb$.
For chargino masses very close to the LEP2 limit, $m_\chi \sim 100$ GeV, one
can have to $A_\chi \gsim 0.5$ inducing contributions to $\Gamma (h
\ra \gamma \gamma$) which exceed the 10\% level. For chargino
masses above 250 GeV, the maximum $A_\chi$ contribution will be below
$\sim 0.2$ for both values of $\tb$, altering the total $h \ra \gamma
\gamma$ width by less than 10\% percent. 

\subsection*{5. Conclusions}

We have analyzed the contribution of charged Higgs bosons, charginos,
sleptons and squark loops to the coupling of the lightest neutral Higgs
boson to two photons in supersymmetric theories. Our aim was to
determine the region of the MSSM parameter space in which one is still
sensitive to the additional SUSY loops, although no SUSY particle has
been produced directly at the LHC or at a future $\ee$ collider with a
c.m.~energy of $\sqrt{s}=500$ GeV. We focussed on the decoupling limit 
where all the additional MSSM Higgs bosons are very heavy. In this limit, 
the $h$ boson will have practically the same properties as the standard 
Higgs particle, and the two--photon decay could be used to discriminate 
between the SM and the MSSM Higgs sectors. Our conclusions are as follows: \s 

The contributions of charged Higgs bosons, sleptons and the scalar
partners of the light quarks including the bottom squarks are extremely
small. This is due to the fact that these particles do not couple to the
Higgs boson proportionally to the mass, and the amplitude is damped by
inverse powers of the heavy mass squared; in addition, the scalar loop
amplitude is much smaller than the dominant $W$ amplitude. For masses
above 250 GeV, the effect of scalar particles [with the exception of the
top squark] on the $h \ra \gamma \gamma$ width does not exceed one 
precent level and can therefore be neglected. \s 

The contribution of the charginos to the two--photon decay width
can exceed the 10\% level for masses close to $m_\chi \sim 100$
GeV, but it becomes smaller with higher masses. The deviation of the
$\Gamma(h \ra \gamma \gamma)$ width from the SM value induced by 
charginos with masses $m_\chi=250$ and $400$ GeV is shown in 
Fig.~8a, as a function of $M_2$ [$\mu$ is fixed by $m_\chi$] for 
$\tb=1.6$ and $50$. For chargino masses above $m_\chi \gsim 250$ 
GeV [i.e. slightly above the limit where charginos can be produced 
at a 500 GeV $\ee$ collider], the deviation is less than $ \sim 8\%$ 
for the entire SUSY parameter space\footnote{The maximum deviation 
is obtained for $M_2$ values slightly above the lightest chargino mass. 
The reason is that the Higgs boson prefers to couple to mixtures of 
gauginos and higgsinos, and in this region the $h \chi_1^+ \chi_1^-$ 
coupling is maximal. For larger $M_2$ values, $\chi_1^+$ is a pure 
higgsino while for smaller $M_2$ values it is a pure gaugino and 
the Higgs coupling is therefore small.}. 
The deviation drops by a factor of two if the chargino mass is 
increased to 400 GeV. \s

Because its coupling to the lightest Higgs boson can be strongly
enhanced, the top squark can generate sizeable contributions to the
two--photon decay width. For stop masses in the $\sim 100$ GeV range, the
contribution could reach the level of the dominant $W$ boson
contribution and the interference is constructive increasing drastically
the decay width. For $\tilde{t}_1$ masses around 250 GeV, the deviation of 
the $h \ra \gamma \gamma$ decay width from the SM value can be still at 
the level of 10\% for a very large off--diagonal entry in the stop mass 
matrix, $m^{LR}_t \gsim 1$ TeV [Fig.~8b]. For larger masses, the deviation 
drops $\sim 1/m_{\tilde{t}_1}^2$ and the effect on the decay width is below
$2\%$ for $m_{\tilde{t}_1} \sim 400$ GeV even at $m^{LR}_t \sim 1$ TeV. 
For small values of $m^{LR}_t$, the deviation does not exceed $-8\%$
even for a light top squark $m_{\tilde{t}_1} \sim 250$ GeV.
For this stop mass value, we have cut out the region 
$m^{LR}_t \lsim 200$ GeV since there, the scalar partners of light 
quarks will have masses smaller than $m_{\tilde{q}} \sim 250$ GeV.
\s 

Thus, the only way to have a contribution to the two--photon 
decay width of the $h$ boson that is larger than 10\% is from
a rather light stop squark $m_{\tilde{t}_1} \lsim 300$ GeV with 
extremely large couplings to the Higgs boson, $m_t^{LR} >1$ TeV. 
However, for $m_{\tilde{t}_1} \sim 300$ GeV and $m^{LR}_t \lsim 1$ TeV,
the scalar partners of the light quarks will have masses around 
$m_{\tilde{q}} \sim 500$ GeV and therefore should be observed at the LHC.
The only way to have a light top squark, $m_{\tilde{t}_1} \sim 300$ GeV, 
while the other squarks are heavier than 1 TeV and escape detection at the
LHC is to increase $m_t^{LR}$ to $\sim 5$ TeV. For such large $m_t^{LR}$
values, the trilinear scalar interaction become extremely strong and 
could lead to dynamically favoured minima of the scalar potential where
charge is not conserved \cite{R26}. [A necessary, though not sufficient 
condition to avoid these false vacua is to choose $m_t^{LR} \lsim 3 
m_{\tilde{q}}$; see Ref. \cite{R26}.] Furthermore, the $h\tilde{t}\tilde{t}$
coupling, $g_{h\tilde{t}\tilde{t}} \sim m_t m^2_{LR}/(2M_Z^2)$,
becomes very large and perturbation theory is endangered. \s

In summary: only chargino and top squark loops can lead to  a sizable 
difference between the two-photon decay width of the lightest SUSY and 
the standard Higgs bosons in the decoupling limit. Charginos with masses 
above the production threshold of a 500 GeV $e^+e^-$ collider will induce 
contributions which are smaller than a few percent in the entire SUSY 
parameter space. Top squarks can induce contributions which exceed the
$10\%$ level only if they are just slightly heavier than $\sim 250$ GeV
and if the off-diagonal entry in the stop mass matrix is very large, 
$m_t^{LR} \gsim 1$ TeV. In the region of parameter space where both 
charginos and top squarks are light, the sum of the two contributions
can exceed the $10\%$ level. \s

One therefore needs a measurement of the Higgs coupling to two photons
at $\gamma \gamma$ colliders with an accuracy better than 10\% to 
discriminate between the standard and the minimal SUSY Higgs scenarios if 
this discrimination could not be achieved in the $\ee$ option of the
collider with a c.m. energy in the 500 GeV range, or if scalar quarks
have not been observed at the LHC. 

\vspace*{1.cm}

\nn {\bf Acknowledgments}: \s

\nn Discussions with Manuel Drees, Howard Haber, Christoph J\"unger,
Janusz Rosiek and Michael Spira are gratefully acknowledged. J.I. is supported 
by a fellowship from the Fundaci\'on Ram\'on Areces and partially by
the Spanish CICYT under contract AEN96-1672. 

\newpage 

\setcounter{equation}{0}
\renewcommand{\theequation}{A.\arabic{equation}}

\subsection*{Appendix: QCD corrections to scalar quark loops}

The calculation of the QCD corrections to scalar quark loop
contributions to the Higgs--two photon coupling can be calculated by
extending low--energy theorems \cite{Hpp} to scalars at the two--loop
level \cite{ggQCD2,LOWE}. For a CP--even Higgs boson $H$ with a mass,
$M_H \ll 2m_{\tilde{Q}}$ which is the case here, these theorems relate
the matrix elements of  the squark contributions to the $H \gamma
\gamma$ vertex to the photon two--point function. In this Appendix, we
will use these theorems to derive the QCD corrections to the scalar
loops due to pure gluon exchange; this part of the correction is
expected to be the dominant one. \s 

Denoting the matrix element of one squark contribution to the photon
self--energy by ${\cal M}_{\tilde{Q}} (\gamma \gamma)$, and the 
corresponding matrix element with an additional light Higgs boson by 
${\cal M}_{\tilde{Q}} (\gamma \gamma H)$, one has at lowest order 
\beq
{\cal M}_{\tilde{Q}}^{\gamma \gamma H} = \left( \sqrt{2}G_F \right)^{1/2} 
\, e_{\tilde{Q}}^2 \, g_{\tilde{Q}}^{H} \, m_{\tilde{Q}} \, \partial 
{\cal M}_{\tilde{Q}}^{\gamma \gamma} / \partial m_{\tilde{Q}} 
\eeq
with $e_{\tilde{Q}}$ and $g_{\tilde{Q}}^H$ the squark charge and its 
reduced coupling to the Higgs boson. To extend this relation to higher 
orders, one has to replace all quantities by their
bare values, differentiate with respect to $m_{\tilde{Q}}$ and then
perform the renormalization. In the case of pure gluon exchange, the
differentiation with respect to the bare squark mass $m_{\tilde{Q}}^0$
can be rewritten in terms of the renormalized mass $m_{\tilde{Q}}$. A
finite contribution to the QCD corrections arises from the anomalous
mass dimension $\gamma_{\tilde{Q}}$ 
\beq
m_{\tilde{Q} }^0 \frac{ \partial} {\partial m_{\tilde{Q}}^0 } = 
\frac{m_{\tilde{Q}}} {1+\gamma_{\tilde{Q}} }~
\frac{ \partial}{\partial m_{\tilde{Q}} }
\eeq
The remaining differentiation with respect to the renormalized squark
mass of the photon two--point function leads to the squark contribution
to the $\beta$ function, $\beta_{\tilde{Q}}$. The final result for the
squark contribution to the Higgs--two--photon coupling can be expressed 
in terms of the effective Lagrangian
\beq
{\cal L}_{\rm eff}^{\tilde{Q}} = \left( \sqrt{2} G_F \right)^{1/2} 
e_{\tilde{Q} }^2 \frac{ g_{\tilde{Q}}^{H} }{4} \frac{\beta_{\tilde{Q}}/
\alpha} {1+\gamma_{\tilde{Q}} } F^{\mu\nu} F_{\mu\nu} H
\eeq
The QCD corrections to the squark loop are then fully determined by the 
anomalous mass dimension of the squarks $\gamma_{\tilde{Q}} = 4\alpha_s/(3 
\pi)$ and by the squark contribution to the $\beta$ function $\beta_{
\tilde{Q}}/ \alpha = 2\alpha/\pi (1+4 \alpha_s/\pi)$ \cite{Jones}. This 
results into a final rescaling of the lowest--order Lagrangian by factor 
\beq
{\cal M}_{\tilde{Q}}^{\gamma \gamma H} \ra {\cal M}_{\tilde{Q}}^{\gamma 
\gamma H} \, \left[ 1 + \frac{8}{3} \frac{\alpha_s}{\pi} \right]
\eeq
This has to be compared with the case of heavy quark loops where the QCD
correction gives a rescaling factor \cite{ggQCD2} 
\beq
{\cal M}_{Q}^{\gamma \gamma H} \ra {\cal M}_{Q}^{\gamma \gamma H}
\left[ 1 - \frac{\alpha_s}{\pi} \right]
\eeq

\newpage

%
\def\npb#1#2#3{{\rm Nucl. Phys. }{\rm B #1} (#2) #3}
\def\plb#1#2#3{{\rm Phys. Lett. }{\rm #1 B} (#2) #3}
\def\prd#1#2#3{{\rm Phys. Rev. }{\rm D #1} (#2) #3}
\def\prl#1#2#3{{\rm Phys. Rev. Lett. }{\rm #1} (#2) #3}
\def\prc#1#2#3{{\rm Phys. Rep. }{\rm C #1} (#2) #3}
\def\pr#1#2#3{{\rm Phys. Rep. }{\rm #1} (#2) #3}
\def\zpc#1#2#3{{\rm Z. Phys. }{\rm C #1} (#2) #3}
\def\nca#1#2#3{{\it Nuevo~Cim.~}{\bf #1A} (#2) #3}
%

\newpage

\subsection*{Figure Captions}

\begin{itemize}

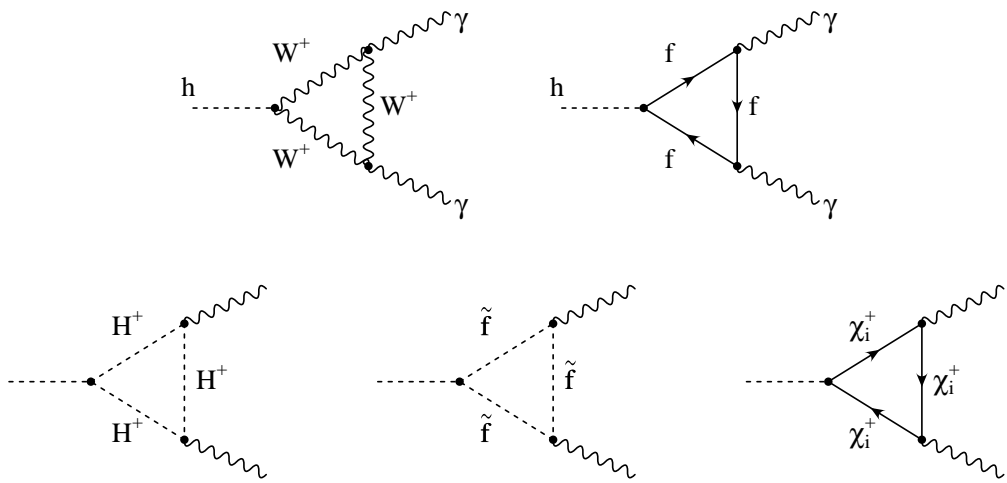
\item[{\bf Fig.~1:}]
Feynman diagrams contributing to the Higgs decay into two photons in 
the MSSM. 

\item[{\bf Fig.~2:}]
The masses of the three MSSM Higgs bosons $h, H$ and $H^\pm$ as a 
function of the pseudoscalar mass $M_A$ for $\tb =1.1, \, 1.6, \, 5$ and 50; 
the common squark mass is set to $m_{\tilde{q}}=250$ GeV (a) and 1 TeV
(b)  and the squark mixing is neglected. 

\item[{\bf Fig.~3:}]
The amplitudes for the contribution of the $W$ boson loop (a) and of the
$t,b$ loops (b) as a function of $M_h$ in the SM [dashed lines] and MSSM
[solid lines]; $M_A$ is fixed to 250 GeV. We have used $m_{\tilde{q}}=250$ 
GeV and neglected squark mixing. 

\item[{\bf Fig.~4:}]
The amplitudes for the contribution of the charged Higgs boson loop (a)
and of the slepton and squark (except stop) loops (b) as functions of
the loop masses for $\tb=1.6, 5$ and 50. We have neglected squark mixing
and for $H^\pm$ and slepton loops we used $m_{\tilde{q}}=250$ GeV; for
sfermion loops we have set $M_A=250$ GeV.  

\item[{\bf Fig.~5:}]
Contours in the ($m^{LR}_t, m_{\tilde{t}_1})$ plane, for which: (a) the
contribution of the stop loops to the $h\gamma \gamma$ coupling is
$|A_{\tilde{t}}| =2, \, 1, \, 0.5$ and $0.2$, and (b) the heavier top
squark mass $m_{\tilde{t}_2}$ and the common mass of the scalar partners
of light quarks $m_{\tilde{q}}$. 

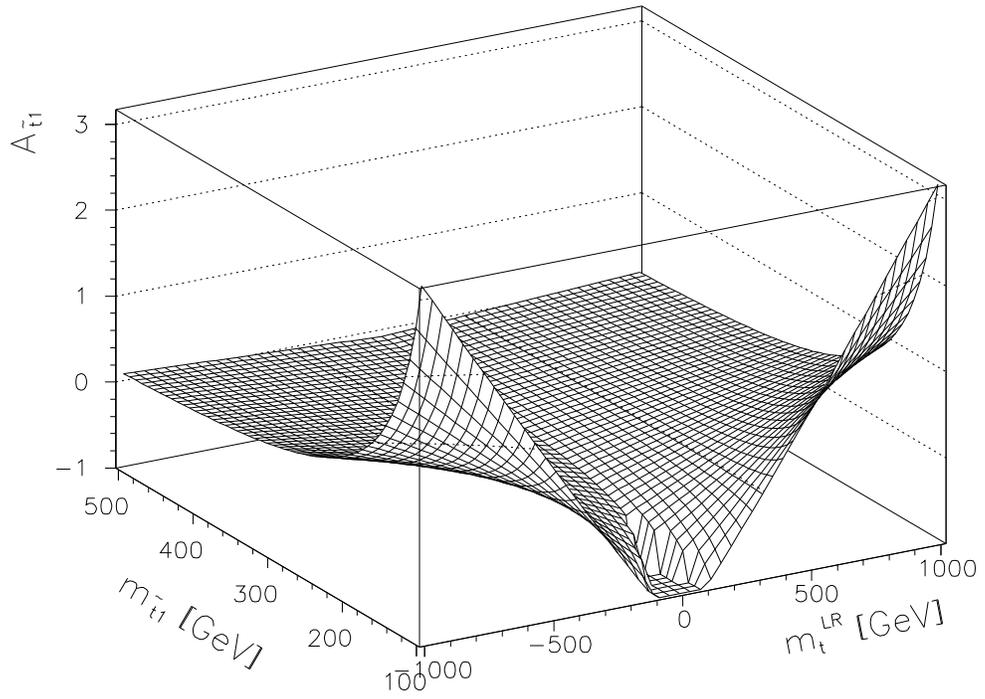
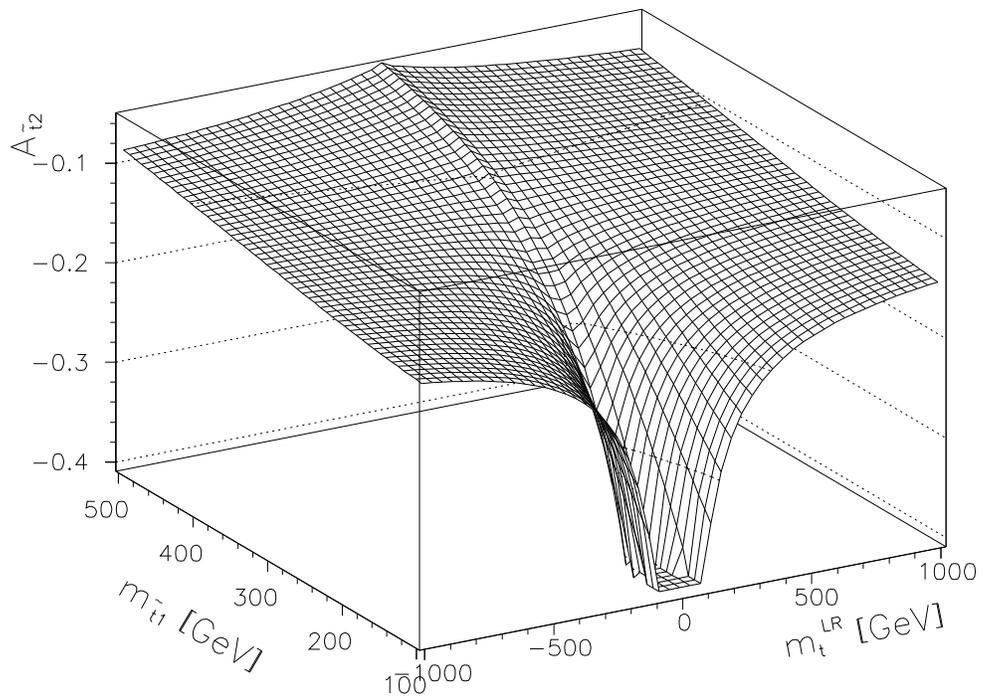
\item[{\bf Fig.~6:}]
The separate contributions of the lightest (a) and the heaviest (b) 
top squark loops to the form factor $A_{\tilde{t}}$ as a function of 
$m^{LR}_t$ and $m_{\tilde{t}_1}$. 

\item[{\bf Fig.~7:}]
Contours in the ($M_2, \mu)$ plane for $\tb=1.6$ (a) and $\tb=50$ (b)
for which the contribution of the chargino loops to the $h\gamma \gamma$
coupling is $|A_{\chi}|=0.1, \, 0.2, \, 0.3$ and $0.5$. Also included 
are the contours for which the lightest chargino mass is $m_{\chi_1^+}
=90$ and $250$ GeV. 

\item[{\bf Fig.~8:}]
The deviations of the SUSY Higgs coupling to two photons from the 
Standard Model value [in \%] for two values of $\tb=1.6$ and 50
and the loops masses $m_i =250$ and $400$ GeV. (a) Deviations 
due to the chargino loops as a function of $M_2$ for both signs of $\mu$,
and (b) deviations due to the top squark loops as a function of 
$m_{t}^{LR}$. 

\end{itemize}

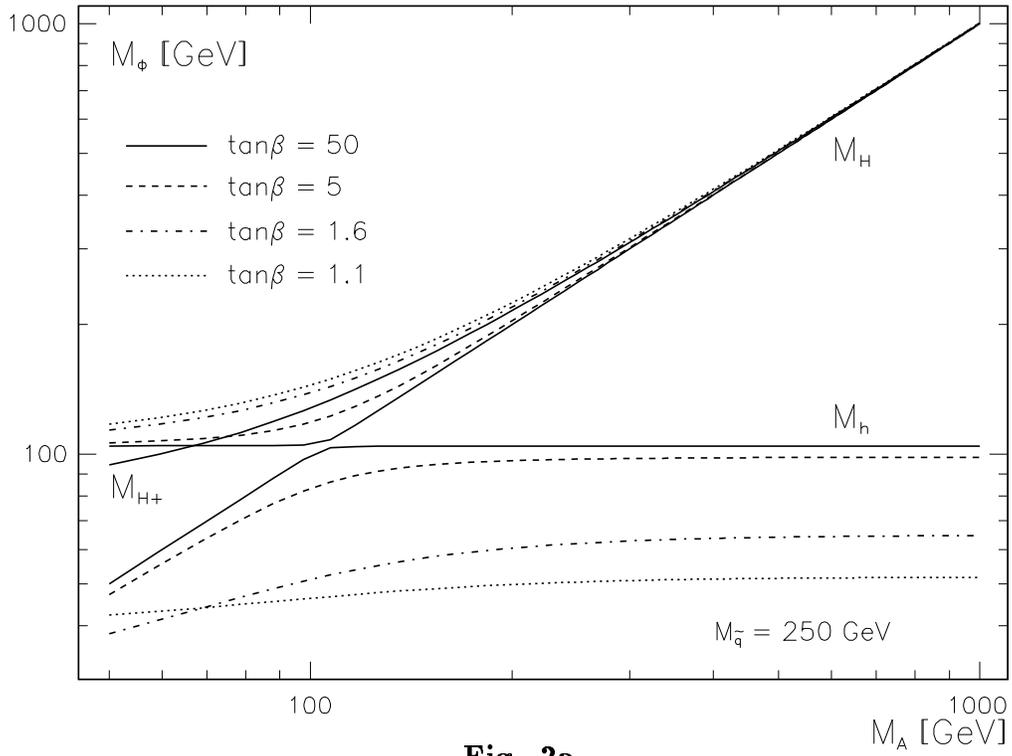
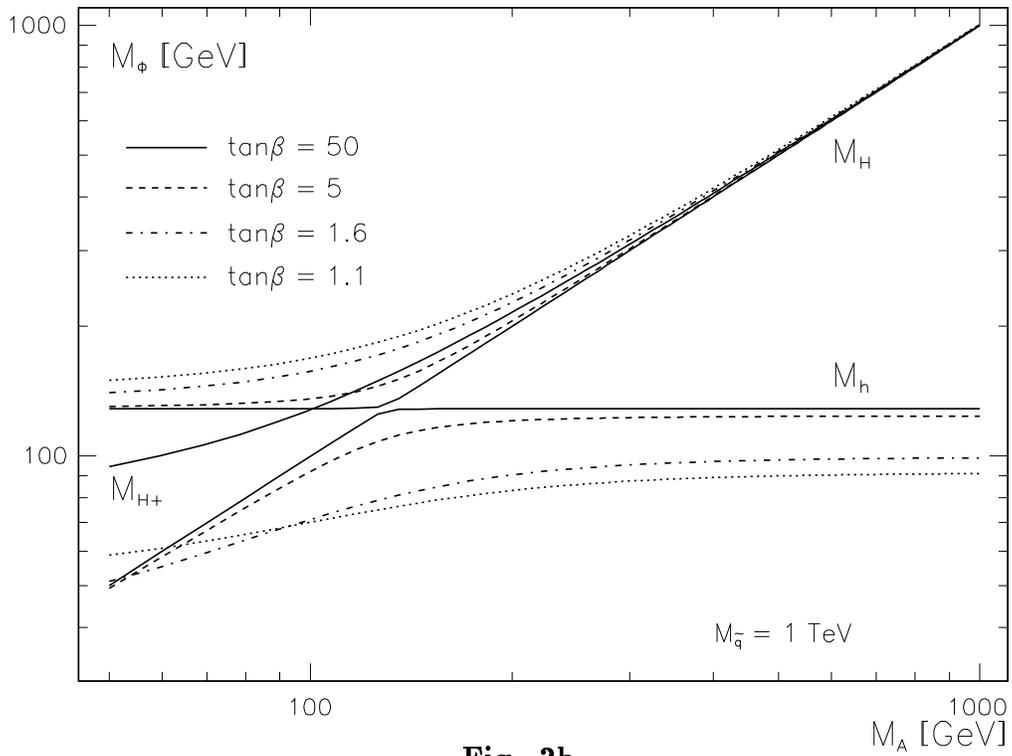
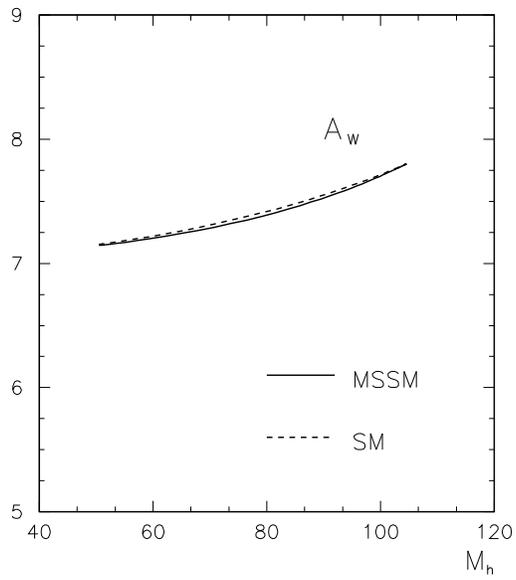
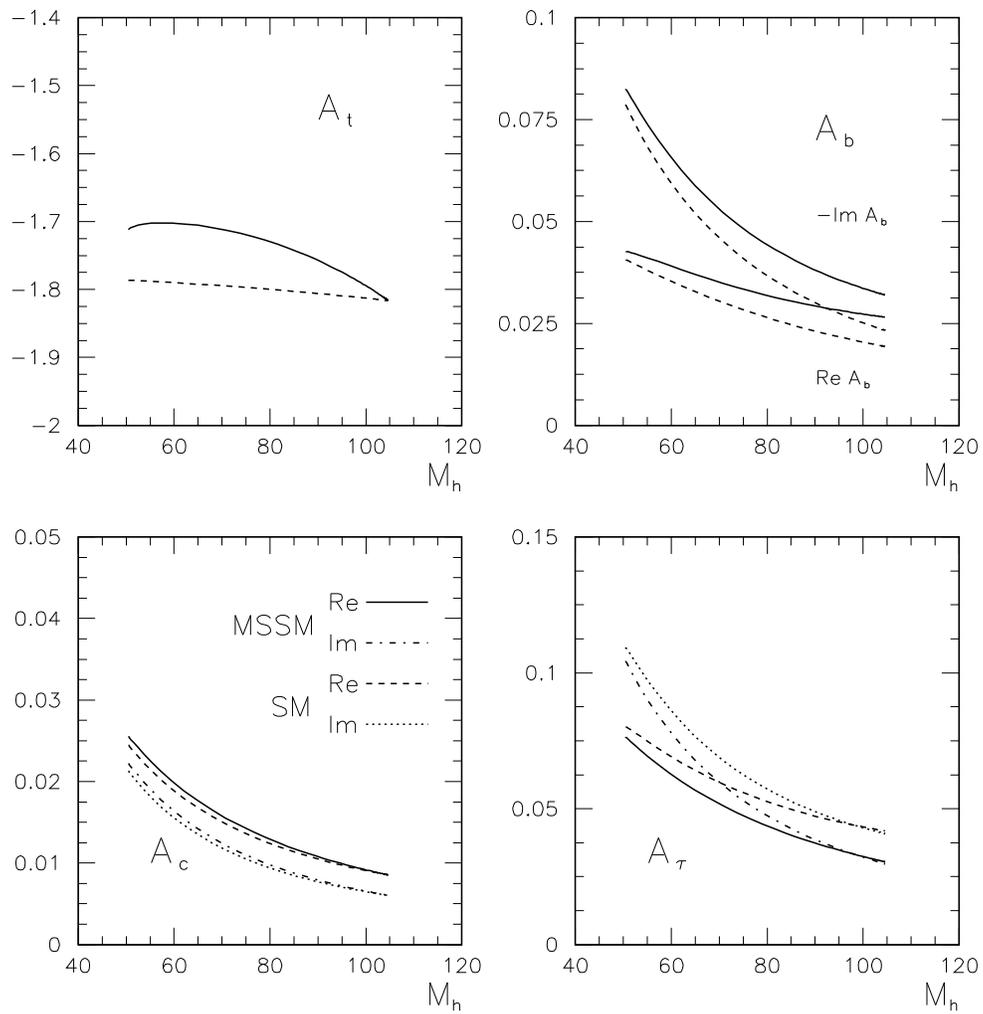
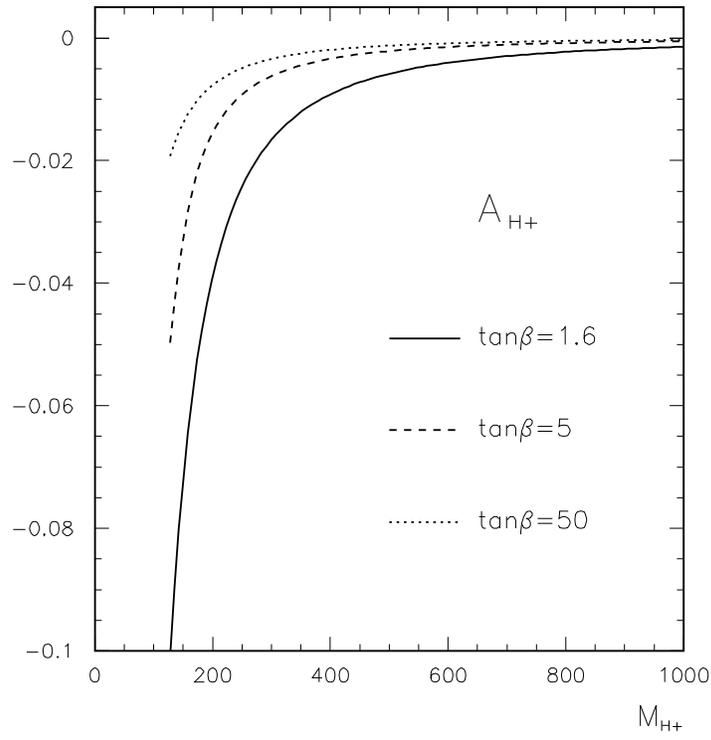
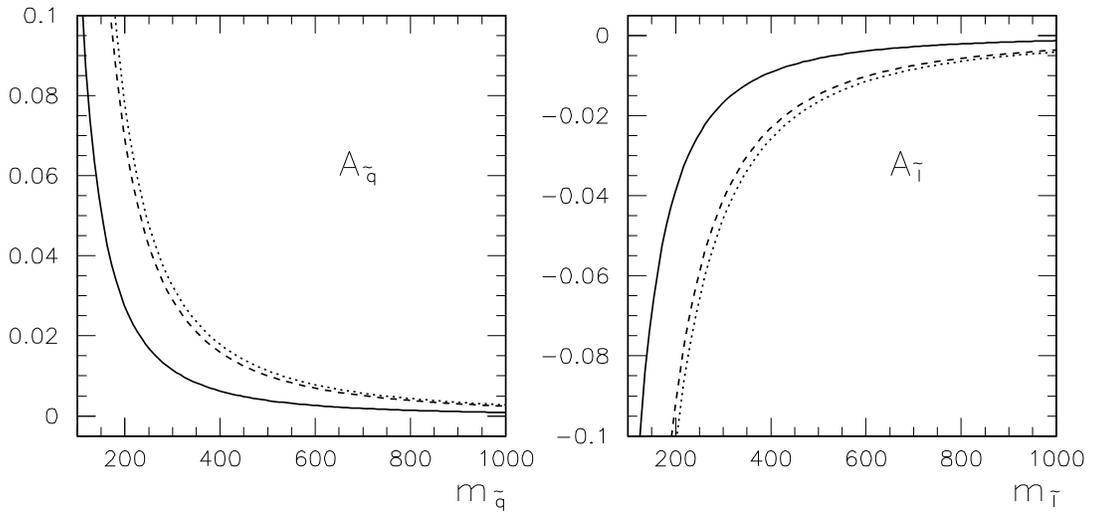
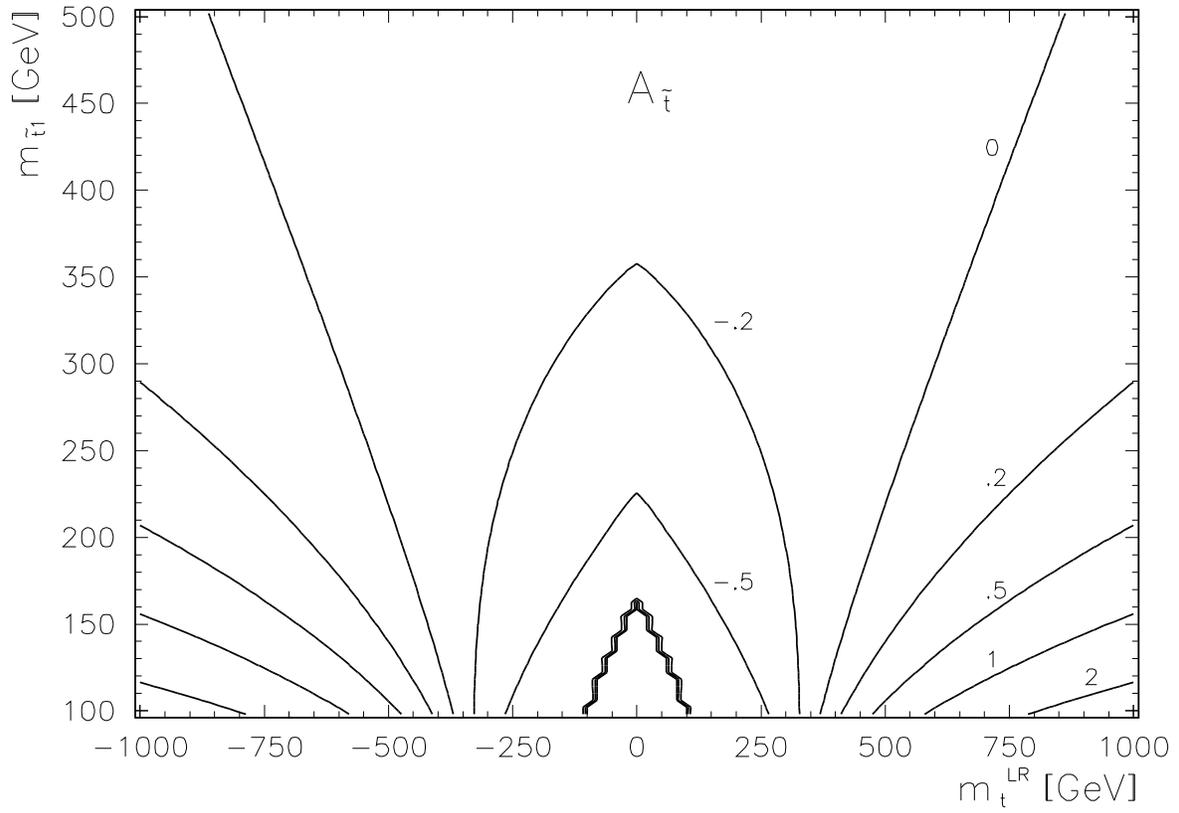
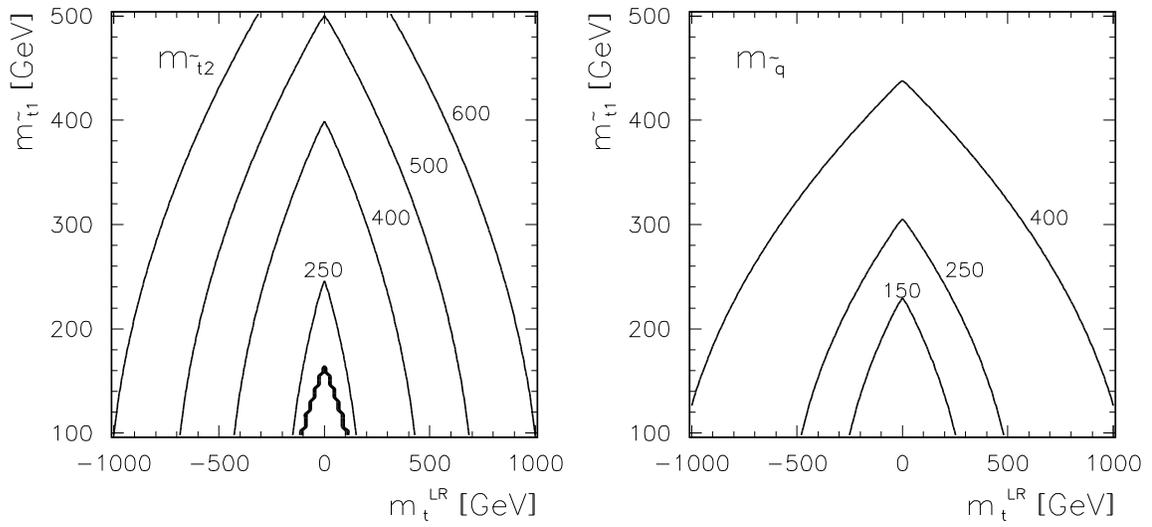
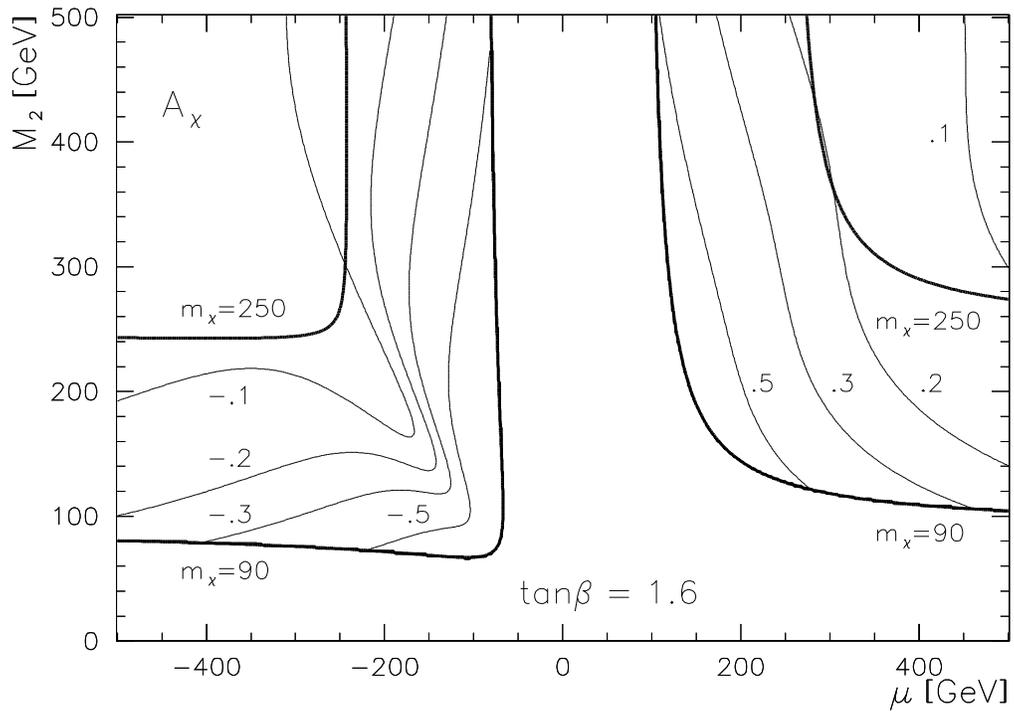
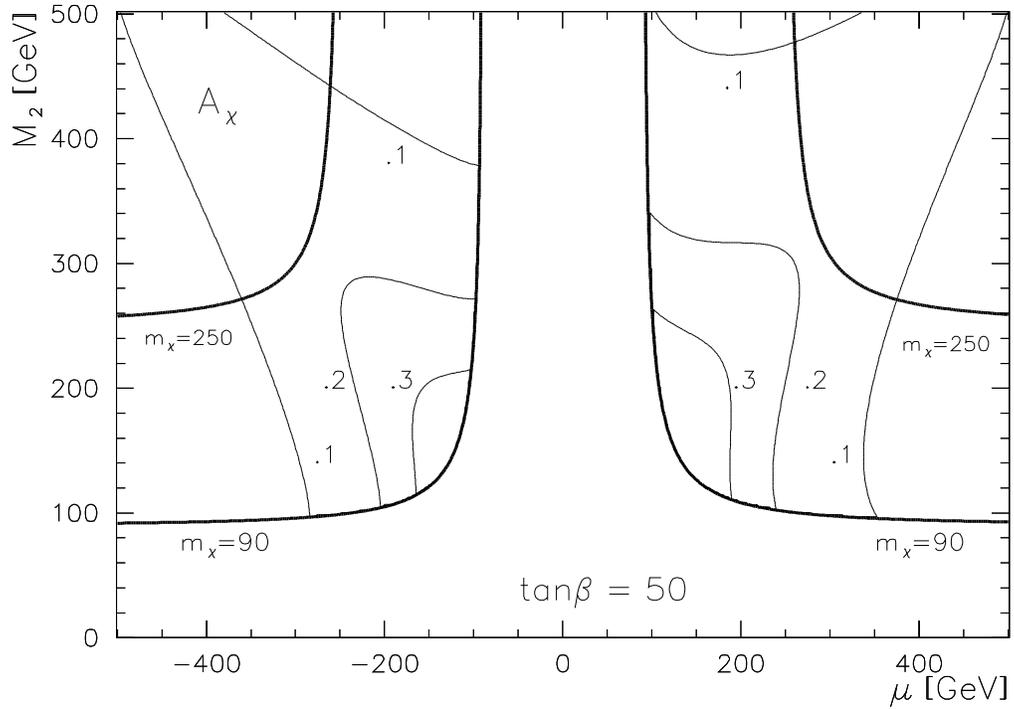
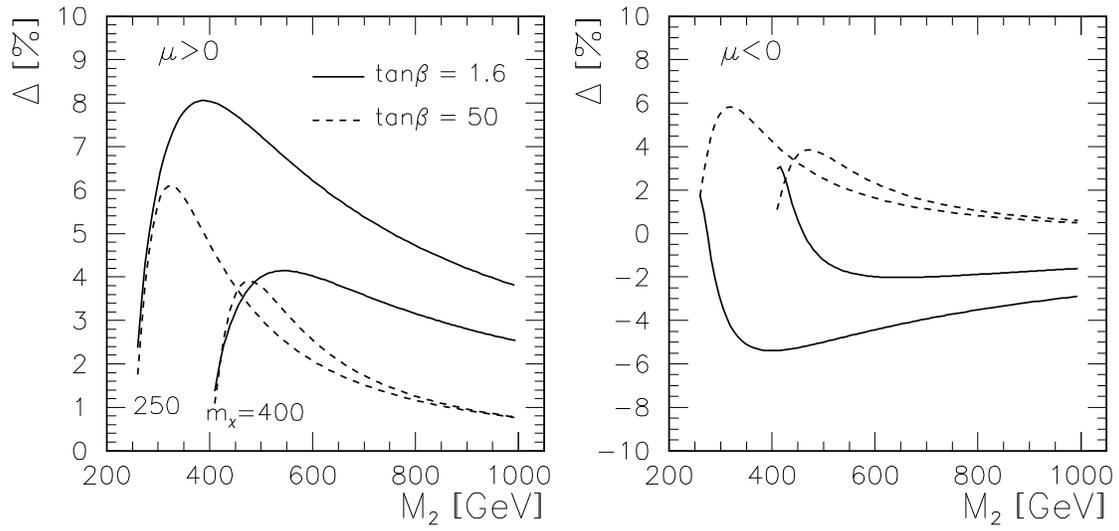
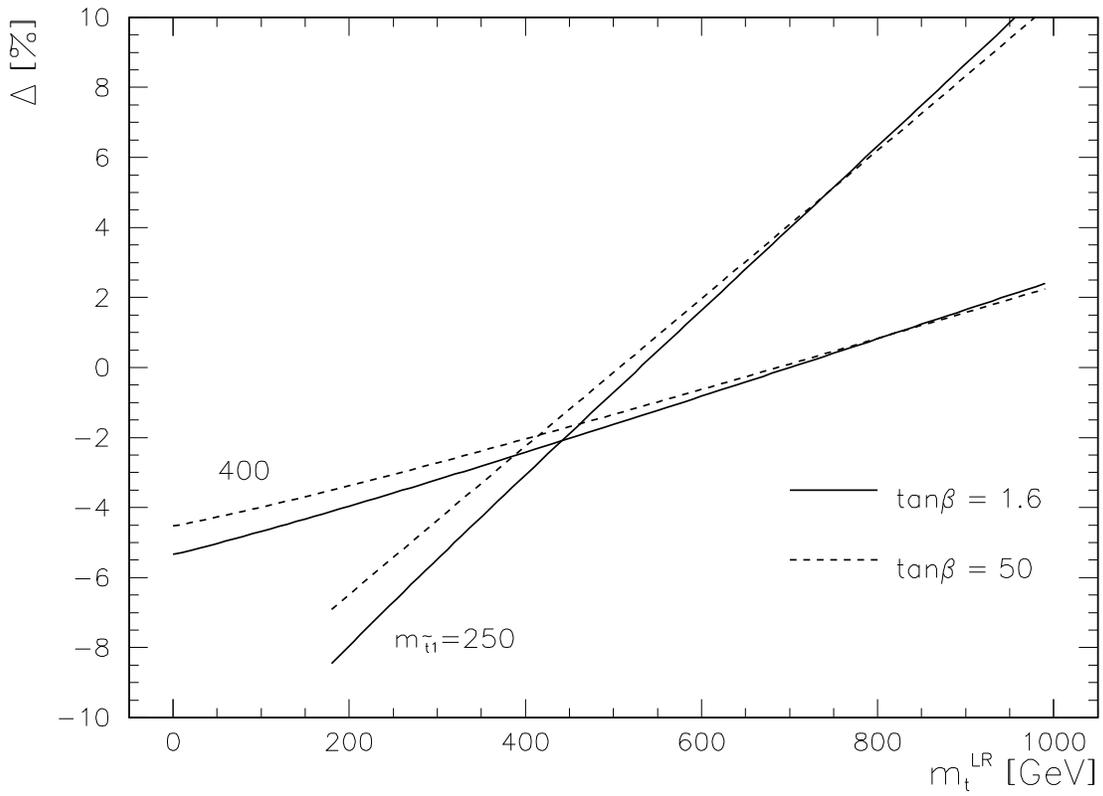
\end{document}